\documentclass[pdftex,twocolumn,epjc3
    ]{svjour3}          

\RequirePackage[T1]{fontenc}

\smartqed  
\RequirePackage{amssymb}
\RequirePackage{mathtools}
\RequirePackage{graphicx}
\RequirePackage{flushend}
\RequirePackage[numbers,sort&compress]{natbib}
\RequirePackage[colorlinks,
citecolor=blue,urlcolor=blue,linkcolor=blue
    ]{hyperref}

\journalname{arXiv:1811.11712}

\RequirePackage{upgreek}
\RequirePackage{braket}
\RequirePackage[inline]{enumitem}

\RequirePackage{subcaption}

\RequirePackage{amsfonts}

\usepackage{soul}

\def\a{\alpha}

\def\e{\varepsilon}

\newcommand\mi{\mathrm{i}} 
\newcommand\me{\mathrm{e}} 
\newcommand\pp{\uppi}      

\newcommand{\dif}{\mathrm{d}}
\newcommand{\Dif}{\mathcal{D}}
\newcommand{\dva}{\updelta\!}

\DeclareMathOperator{\sech}{sech}

\DeclareMathOperator{\arcsinh}{arsinh}

\DeclareMathOperator{\sd}{sd}
\DeclareMathOperator{\cd}{cd}

\DeclareMathOperator{\am}{am}

\newcommand{\rbr}[1]{{\left(#1\right)}}
\newcommand{\sbr}[1]{{\left[#1\right]}}
\newcommand{\cbr}[1]{{\left\{#1\right\}}}

\newcommand{\vbr}[1]{{\left|#1\right|}}

\newcommand{\rfun}[2]{{#1}\mathopen{}\left(#2\right)\mathclose{}}
\newcommand{\sfun}[2]{{#1}\mathopen{}\left[#2\right]\mathclose{}}

\def\a{\alpha}

\def\e{\varepsilon}

\makeatletter
\newcommand*{\rom}[1]{\expandafter\@slowromancap\romannumeral #1@}
\makeatother

\begin{document}

\title{Dynamically Assisted Schwinger Effect at Strong Coupling with Its Holographic Extension}

\author{Chen Lan\thanksref{e1,addr1}
        \and
        Yi-Fan Wang\thanksref{e2,addr2}
        \and
        Huifang Geng\thanksref{e3,addr1}
        \and
        Alexander Andreev\thanksref{e4,addr1,addr3}
}

\thankstext{e1}{stlanchen@yandex.ru}
\thankstext{e2}{yfwang@thp.uni-koeln.de}
\thankstext{e3}{Huifang.Geng@eli-alps.hu}
\thankstext{e4}{Alexander.Andreev@eli-alps.hu}

\institute{ELI-ALPS Research Institute, ELI-Hu NKft, Dugonics t\'er 13, 6720 Szeged, Hungary\label{addr1}
          \and
          Institut f\"ur Theoretische Physik, Universit\"at zu K\"oln, Z\"ulpicher Stra\ss e 77, 50937 K\"oln, Germany\label{addr2}
          \and
          Saint-Petersburg State University, Ulyanovskaya str.\ 1, Petrodvorets, Saint-Petersburg 198504, Russia\label{addr3}
}

\date{Last update: June 21, 2019}

\maketitle

\begin{abstract}
At strong-coupling and weak-field limit, the scalar Schwinger effect 
is studied by the field-theoretical method of worldline instantons
for dynamic fields of single-pulse and sinusoidal types. 
By examining the Wilson loop along the closed instanton path,
corrections to the results obtained from weak-coupling approximations are discovered.
They show that this part of contribution for production rate becomes dominant as Keldysh parameter increases, it makes the consideration at strong coupling turn out to be indispensable for dynamic fields. 
Moreover a breaking of weak-field condition similar to constant field also happens around the critical field, defined as a point of vacuum cascade.
In order to make certain whether the vacuum cascade occurs beyond the weak-field condition, following Semenoff and Zarembo's proposal, 
the Schwinger effects of dynamic fields are studied 
with an $\mathcal{N}=4$ supersymmetric Yang--Mills theory in the Coulomb phase.  
With the help of the gauge/gravity duality, the vacuum decay rate is evaluated 
by the string action with instanton worldline as boundary,
which is located on a probe D3--brane.
The corresponding classical worldsheets are estimated by perturbing the integrable case of a constant field.
\end{abstract}

\section{Introduction}
\label{sec:intro}

The Extreme Light Infrastructure (ELI) is designed to produce 
the highest power and intense laser worldwide~\cite{mourou2011,dunne2009}. 
It has a potential of reaching ultra-relativistic intensities, 
challenging the \emph{Schwinger limit}
$E_\text{s} \coloneqq m^2/e 
\approx 1.32\times 10^{18} \mathrm{V\, m^{-1}}$. 
As laser field approaches this value, 
vacuum becomes unstable, 
and a large amount of charged particles produces in pairs,
so that laser loses the energy, and its intensity stays within the upper limit.
However, Schwinger effect is not only a phenomenon in electromagnetism, 
but a universal aspect of quantum vacuum in the presence of a $\mathsf{U}(1)$ gauge field 
with a classical background, see e.g.\ \cite{wong1994,gorsky2002,nayak2005a,nayak2005b,semenoff2011}. 

The pair-production rate in the constant electric field had been
pioneered by Sauter, Heisenberg and Euler~\cite{sauter1931,heisenberg1936}, 
and the corresponding \emph{Effect} was named after Schwinger~\cite{schwinger1951}, 
who did the calculation based on field-theoretical approaches,
see e.g.\ \cite{gelis2016,ruffini2010} for a current review.
A semiclassical approach called worldline instantons
has been introduced more recently to the study of constant and inhomogeneous fields 
in the small-coupling and weak-field approximations~\cite{affleck1982,dunne2005b},
where the production rate, in Wick-rotated Euclidean space, 
is represented by a worldline path integral.
The so-called \emph{worldline instantons} are the periodic saddle points relevant for
a calculation of integral by the steepest descent method. 
The extension of inhomogeneous fields originates from more practical purpose. As analysed in \cite{dunne2005b}, $1$-D dynamic electric fields reduce the critical value of Schwinger effect, such 
that pair production from vacuum is more close to experimentally observable conditions.

The Schwinger effect at arbitrary coupling in constant field 
has also been studied in \cite{affleck1982} at the weak-field condition, which is considered originally in order to overcome some obstacles from direct application of Schwinger's approach. However the later observation (e.g. \cite{kawai2015}) found that the weak-field condition is broken around the critical field, defined as a point of vacuum cascade, such that the mechanism in weak-field condition loses its prediction for this vacuum phenomenon. Inspired by the similar existence of critical value of electric field in string theory, it is of a possibility to clarify the vacuum cascade in the Coulomb phase beyond the weak-field condition  
with the help of gauge/gravity
duality~\cite{semenoff2011,sato2013a,sato2013c}. This is
also known as the Semenoff--Zarembo construction,
where the production rate has been obtained by calculating
the classical action of a bosonic string, 
which is attached to a probe D3--brane and coupled to a Kalb--Ramond field.

Since the instanton action in the production rate is equivalent to
the string action, which is proportional to the area, calculation of
the rate is related to integrating the classical equations of
motion for bosonic string in the given external field.
In other words, duality converts the problem to evaluating the area of a
\emph{minimal surface}~\cite{kruczenski2014,dekel2015} in Euclidean $\mathrm{AdS}_3$, 
the boundary of which is assumed to be
the trajectory of the worldline on the probe D3--brane.
In mathematics, such a Dirichlet problem is known as the \emph{Plateau's problem}~\cite{Harrison2016}.

In this work, we consider the scalar pair production in a 
dynamic external field of single-pulse and sinusoidal types
at strong coupling and weak-field limit based on the method of worldline instantons, 
which is explained in sections \ref{sec:arbcoup} and \ref{sec:wilson}.
We first show by a field-theoretical approach, 
that besides the enhancement due to the dynamics of electric fields, 
a further contribution to the production rate arises from
the Wilson loop, and it becomes dominant in production rate as Keldysh parameter increases. However, such a correction seems to 
diverge as Keldysh adiabaticity parameter $\gamma \to \infty$ based on our estimated formula; and it leads to a contradiction to the weak-field condition,
so that near the critical field, the method itself breaks down, 
and the prediction of a vacuum cascade becomes unclear.
To overcome the problem of breaking weak-field condition, we then follow Semenoff and Zarembo's proposal
in sec.\ \ref{sec:duality}, applying the gauge/gravity duality to the
Schwinger effect in the Coulomb phase of an $\mathcal{N}=4$ supersymmetric
Yang--Mills theory.
The classical solution of the corresponding string worldsheets are
estimated by perturbing the solvable case of a constant external electric field.

\section{Worldline instantons at strong coupling}
\label{sec:arbcoup}

The worldline instanton approach is a semiclassical calculation 
realised by the worldline path integral representations~\cite{schubert2001,schubert1996},
in which the so-called worldline instanton is a periodic solution of the stationary
phase in the path integral.
Based on this method, the pair-production rate $\Gamma$ for a massive scalar QED 
in the small-coupling and weak-field approximation is computed
by~\cite{affleck1982,dunne2005b,kawai2015}
\begin{equation}\label{eq:wordlineinstanton}
V_4 \Gamma = - 2{\mathrm{Im}} \int \Dif x\, \frac{1}{m}\sqrt{\frac{2\pp}{T_0}}\,
\rfun{\exp}{-S_{\text{inst}}},
\end{equation}
where $V_4$ is the 4-volume, $T_0$ is a constant, given by
\begin{equation}
    T_0=\frac{\sqrt{\int^1_0 \dif\tau\, \dot x^2}}{m},
\end{equation}
and $S_{\text{inst}}$ represents the action of worldline instanton, i.e.\
\begin{equation}
\label{eq:wordlineaction}
S_{\text{inst}} = m^2 \sqrt{\int^1_0 \dif\tau\, \dot x^2}- \mi e \int^1_0 \dif \tau\, 
B \cdot \dot x.
\end{equation}
In eq.\ \eqref{eq:wordlineaction},
$B_\mu$
is a classical background field, not to
be confused with the fluctuation part of the $\mathsf{U}(1)$ gauge field 
$A_\mu$, 
which is also included in the initial setup of path integral, but cancelled due to considerations at small coupling.

The path integral in eq.\ \eqref{eq:wordlineinstanton} can be computed 
by the stationary phase approximation in \emph{weak-field condition}
\begin{equation}
\label{eq:weak-field}
    m\sqrt{\int^1_0\dif u\;\dot x^2}\gg 1
\end{equation}
For more general cases with 1D temporally inhomogeneous fields $B_1(x_0)$, 
the worldline instanton is obtained by solving the instanton equation
with periodic boundary condition
\begin{equation}
\label{eq:instantoneq}
    x'_1(x_0)=\pm \frac{\mi e }{m}\frac{B_1(x_0)}{\sqrt{1+\left(
    \frac{e B_1(x_0)}{m}
    \right)^2}}.
\end{equation}
Thus the pair-production rate can be written as
\begin{equation}
\label{eq:decayrate}
\Gamma \sim \rfun{\exp}{-\frac{\pp m^2}{e E} f(\gamma)},
\end{equation}
where $f(\gamma)$ is a monotonically decreasing 
function%
of Keldysh parameter \cite{keldysh1965}
$\gamma=m \omega/(e E)$ and calculated by substituting the solution of
eq.\ \eqref{eq:instantoneq} into the worldline action with the corresponding
background field $B_\mu$. 
In addition, the weak-field condition eq.\ \eqref{eq:weak-field} is of various forms according to different dynamic fields, because it depends on specific instanton solution. 

At arbitrary coupling constant but with weak-field condition \cite{affleck1982}, the production rate is modified by a factor,
\begin{equation}
V_4 \Gamma = - 2{\mathrm{Im}} \int \Dif x\,
\frac{1}{m}\sqrt{\frac{2\pp}{T_0}}\, \rfun{\exp}{-S_{\text{inst}}}
\Braket{W},
\end{equation}
where $\Braket{W}$ is the
average of $\mathsf{U}(1)$ Wilson loop $W$
\begin{equation}
W=\rfun{\exp}{\mi e \oint_C A\cdot \dif x}.
\end{equation}
The integral path $C$ is along the 2D trajectory of worldline instanton.
In the Feynman gauge of the $\mathsf{U}(1)$ field, the Wilson loop
becomes~\cite{srednicki2007}
\begin{equation}
\label{eq:wilsonloop}
\braket{W}=\rfun{\exp}{
	-\frac{e^2}{8\pp^2}
	\mathcal{A}}.
\end{equation}
After simplification, 
$\mathcal{A}$ can be represented via a double contour integrals
\begin{equation}\label{eq:aint}
	\mathcal{A}=\oint_C  \oint_C
	    \frac{\dif  x\cdot  \dif y}{(x-y)^2}.
\end{equation}
For a constant electric field, the worldline instanton is a circle, 
and the production rate of a single scalar pair is given by 
\begin{equation}
\Gamma=\frac{(e E)^2}{(2\pp)^2} \me^{-\pp\frac{ E_\text{s}}{E}},
\qquad
E_\text{s} \coloneqq \frac{m^2}{e},
\label{eq:schwingerlimit}
\end{equation}
where $E_\text{s}$ is the Schwinger limit in natural units.
Consideration at arbitrary coupling~\cite{affleck1982} leads to a correction from eq.\ \eqref{eq:wilsonloop}
\begin{equation}
\Gamma=\frac{(e E)^2}{(2\pp)^2} \me^{-\pp\frac{ E_\text{s}}{E}+\frac{e^2}{4}},
\end{equation}
see also sec.\ \ref{subsec:constant}. 
Our aim in the next section is to calculate the Wilson loops in
eq.\ \eqref{eq:wilsonloop} along two specific worldlines separately. 
The first one is the worldline instanton in a single-pulse field, 
whereas the second is in a sinusoidal field. 
We show that the corrections due to the dynamic fields depend on Keldysh parameter, and
the weak-field condition is also broken in these two cases, 
as noted in e.g.\ \cite{kawai2015}.

\section{Wilson loops along worldline instanton paths}
\label{sec:wilson}

The Wilson loop in eq.\ \eqref{eq:wilsonloop} plays a central role in the case 
of arbitrary coupling, the integral eq.\ \eqref{eq:aint} standing on the 
exponent diverges as $x$ approaches $y$. 
A regulator $\varepsilon$ has been introduced in \cite{polyakov1980},
such that eq.\ \eqref{eq:wilsonloop} becomes
\begin{equation}
\label{eq:wilsonphase}
\mathcal{A}_\e = \int_{-\pp}^{\pp} \dif s\,\int_{-\pp}^{\pp}
\dif t\, \frac{ x'(s) \cdot x'(s+t) }{[x(s+t)-x(s)]^2+\e^2},
\end{equation}
where prime indicates derivative with respect to the instanton parameter,
and $s$ and $t$ can be understood as angular coordinates for the instanton. 

One sees that the integrand in $\mathcal{A}_0$ behaves like $t^{-2}$ as $t \to 
0$, rendering the integral divergent. Introducing $\varepsilon$ makes the 
integral regular, and the divergence can now seen explicitly by expanding the 
numerator and denominator of the integrand separately. Up to $\rfun{O}{t^2}$,
the numerator reads
\begin{align}
&\quad\,
x'(s)\cdot  x'(s+t)
\nonumber \\
&= x'(s)^2 +
x'(s) \cdot  x''(s) t + \frac{1}{2}  x'(s) \cdot  x'''(s) t^2+O(t^3)
\nonumber \\
&= x'(s)^2+\frac{1}{2}  x'(s) \cdot  x'''(s) t^2+O(t^3),
\label{eq:arbcoup-num-1}
\end{align}
whereas the denominator becomes
\begin{equation}
\sbr{ x(s+t)-  x(s)}^2+\e^2
= \e^2 + \rfun{x'}{s}^2 t^2 +  O(t^4).
\end{equation}
The condition $x'(s) \cdot x''(s) \equiv 0$ has also been used, because 
all worldline instantons satisfy $\rfun{x'}{s}^2 =a^2$, where
$a$ is defined in the same way as in \cite{dunne2005b}.
Hence up to second order of $t$, one has
\begin{align}
    {}^2\!\mathcal{A}^{(0)}_\e
    &=\int^{\pp}_{-\pp}\dif s \, x'(s)^2
    \int^{\pp}_{-\pp}\frac{\dif t}{x'(s)^2 t^2+\e^2}
\nonumber \\
    & =\frac{\pp}{\e} \int^{\pp}_{-\pp}\dif s\,\sqrt{x'(s)^2}-4+O(\e),
\label{eq:zeroorder}
\end{align}
    ${}^2\!\mathcal{A}^{(1)}_\e \equiv 0$,
and
\begin{align}
   {}^2\! \mathcal{A}^{(2)}_\e &=
    \frac{1}{2}\int^{\pp}_{-\pp}\dif s\,
     x'(s)\cdot  x'''(s)
    \int^{\pp}_{-\pp}
    \frac{t^2\,\dif t}{ x'(s)^2 t^2+\e^2}
    \nonumber \\
    &=\pp \int^{\pp}_{-\pp}\dif s\,
    \frac{ x'(s)\cdot x'''(s)}{ x'(s)^2}+O(\e).
    \label{eq:secondorder}
\end{align}
One sees that the divergence of $\mathcal{A}$ now is 
removed by introducing a \emph{subtraction term}
\begin{equation}
\label{eq:perimeterlaw}
\dva\mathcal{A}=\frac{\pp}{\e}\int_{-\pp}^{\pp} \dif s \sqrt{\rfun{x'}{s}^2},
\end{equation}
which is an example of the \emph{perimeter law}, depicting the behaviour
of the Wilson loop in Euclidean space~\cite[ch.\ 82]{srednicki2007}. Since 
$\rfun{x'}{s}^2 =a^2$ is independent of integration variable, the subtraction
term can be
worked out as
\begin{equation}
\label{eq:regulator}
\dva\mathcal{A}=\frac{2 \pp^2 a}{\e}.
\end{equation}
After subtracting this term from eq.\ \eqref{eq:wilsonphase},
one obtains the physical $\mathcal{A}_{\text{phy}}$ by taking
the limit of regularised $\mathcal{A}_{\text{reg}}$
\begin{equation}
\mathcal{A}_{\text{phy}} \coloneqq 
\lim_{\e\to 0}\mathcal{A}_{\text{reg}}
\coloneqq \lim_{\e\to 0}\rbr{\mathcal{A}_\e-\dva\mathcal{A}}.
\end{equation}
The physical Wilson loop is then given by
\begin{equation}
    \Braket{W}_{\text{phy}}=
    \rfun{\exp}{-\frac{e^2}{8\pp^2} \mathcal{A}_{\text{phy}}},
\end{equation}
which appears as a factor in the final expression of the production rate.

In our practice with the dynamic fields, $\mathcal{A}_\varepsilon$ 
has yet to be worked out in a closed form, and its estimation is to be 
discussed in sec.\ \ref{ssec:est-A}.

\subsection{Estimation of $\mathcal{A}_\varepsilon$ and $\mathcal{A}_\text{phy}$}
\label{ssec:est-A}

To derive an approximation of $\mathcal{A}_\varepsilon$, one may keep the finite
terms in eq.\ \eqref{eq:zeroorder} and  \eqref{eq:secondorder}, i.e.\ 
\begin{equation}
\label{eq:texpansion}
   {}^2\! \mathcal{A}_{\text{reg}}\approx
  -4+ \pp \int^{\pp}_{-\pp}\dif s\,
    \sbr{\frac{ x'(s)\cdot x'''(s)}{ x'(s)^2}}+O(\e)
\end{equation}
This will be called a \emph{$t$-expansion} up to second order of $t$,
which follows straightforwardly from the 
separation of divergent term. 
This method belongs to rational approximation.
Furthermore in our application in dynamic fields,
this expansion can be worked out in a closed form easily. Also note the $-4$ 
term in eq.\ \eqref{eq:texpansion}, which will be mentioned again later with
example.
The validity of $t$-expansion is closely related to the uniform convergence of integrand, and demonstrated in \ref{app:texpansion}. If one uses diagonal Pad\'e approximant for integrand rather than $t$-expansion, the convergence is obvious 
\cite{nuttall1970,zinn1974,gonchar1983}.

Alternatively, one may also expand the integrand of $\mathcal{A}_\epsilon$
with respect to $\gamma$ for the well-defined point $\gamma_0$, i.e.
\begin{equation}
\mathcal{A}_{\e}=
\int^{\pp}_{-\pp}\dif s\int^{\pp}_{-\pp}\dif t\, \left[\sum_n f_n(\e)
\rbr{\gamma-\gamma_0}^n\right]
\end{equation}
where $n\in \mathbb{Z}$. 
This will be called a $\gamma$-expansion.
If the sequence $\{f_n(\e)\}$ is integrable term by term, such that the interchanging summation and integration is valid, then comparing with the similar expansion of $\dva\mathcal{A}$, one obtains the finite terms of each order by
\begin{equation}
   \mathcal{ A}^{(n)}_{\text{reg}} =
    \frac{1}{\e}\sum_{k=0} F_{n,k}\e^k-\frac{2\pp^2 a_n}{\e}
\end{equation}
where $F_{n,k}$ are coefficients of expansion for $F_n$ with respect to $\e$
\begin{equation}
    F_n = \int^{\pp}_{-\pp}\dif s\int^{\pp}_{-\pp}\dif t\,f_n(\e)
\end{equation}
while $a_n$ are coefficients of expansion for $a(\gamma)$ with respect to $\gamma$. Since $\e=0$ is the pole of first order for $A_\e$ as a function of $\e$, $F_n$ having the same pole of $\e$ is obviouse. For those points, where interchanging operation not long stands, this approach fails, e.g. at $\gamma_0\to+\infty$.
In our application in dynamic fields, this expansion can also be worked out
in a closed form at each order. However, the number of terms in the expansion increases 
exponentially, and the result are obtained by computer algebra system.

Yet another way of estimating $\mathcal{A}_\text{phy}$ is numerical 
integration, in which the regulator $\varepsilon$ is still needed. There are 
polynomial contributions of $\varepsilon$ in the bare term 
$\mathcal{A}_\varepsilon$, and one might think taking a small $\varepsilon$ 
would give a good result. However, $\mathcal{A}_\varepsilon$ and the counter 
term $\dva\mathcal{A}$ both diverges like $\varepsilon^{-1}$ as 
$\varepsilon\to 0$. A small $\varepsilon$ leads to a numerically dissatisfying 
operation, in which two big numbers cancels, yielding a small result and a great
loss of significance. This problem becomes catastrophic for the dynamic fields
when $\gamma \to 0^+$.
In order to overcome the potentially catastrophic cancellation, we use linear
extrapolation near $\varepsilon = 0$, in which for each $\gamma$, 
$\mathcal{A}_\text{reg}$ is numerically calculated for 
several different values of $\varepsilon$. The limit of $\varepsilon \to 0$
is then obtained by linearly extrapolate the series of results with respect
to $\varepsilon$. In this approach, the error due to extrapolation can also
be obtained by estimation of the parameters in linear regression.
Furthermore, in our application the numerator and denominator in eq.\
\eqref{eq:wilsonphase} scales as $\gamma^{-2}$ when $\gamma \to +\infty$,
so that for a fixed $\varepsilon$, at large $\gamma$ the regularised integrand
is dominated by the regulator on the denominator. This is overcome by
scaling $\varepsilon$ accordingly, such that the subtraction term in
eq.\ \eqref{eq:wilsonphase} remains constant with respect to $\gamma$.

\subsection{Constant electric field}
\label{subsec:constant}

For a constant electric field, the worldline instanton is a circle
of radius $R=m/(e E)$, given by \cite{affleck1982,dunne2005b}
\begin{equation}
x_0=R \sin(u),
\quad
x_1=R \cos(u),
\qquad
u\in[-\pp, \pp],
\end{equation}
where the zeroth component $x_0$ denotes the Euclidean time.
The instanton action for single pair production is obtained by
substituting this solution into eq.\ \eqref{eq:wordlineaction},
yielding
\begin{equation}
S_0 = \frac{\pp m^2}{e E}.
\end{equation}
The weak-field condition $m\sqrt{\int^1_0 \dif u\,\dot x^2}\gg1$ \cite{dunne2005b} implies
\begin{equation}
\label{eq:weakcondition}
E\ll 2\pp E_\text{s},
\end{equation}
where $E_\text{s}$ is defined in eq.\ \eqref{eq:schwingerlimit}.

On the other hand, the integral in eq.\ \eqref{eq:wilsonphase} and the
perimeter law in eq.\ \eqref{eq:perimeterlaw} can be evaluated explicitly as
\begin{equation}
\mathcal{A}=2 \pp ^2 \rbr{\frac{2 R^2+\e ^2}{\e  \sqrt{4 R^2+\e^2}}-1},
\qquad
\label{eq:substract}
\dva \mathcal{A}=-\frac{2 \pp ^2 R}{\e }.
\end{equation}
Hence the regularised Wilson loop for a single scalar pair
can be recovered as \cite{affleck1982}
\begin{equation}
\label{eq:wlconstant}
\braket{W}_{\text{reg}}=\rfun{\exp}{\frac{e^2}{4}},
\end{equation}
and the production rate is given by
\begin{equation}
\Gamma \sim \rfun{\exp}{-\frac{\pp m^2}{e E} + \frac{e^2}{4}},
\end{equation}
from which the critical field for vacuum cascade can be estimated by
\begin{equation}
E_\text{c} = \frac{E_\text{s}}{\a} \approx 137 E_\text{s},
\qquad
\a \coloneqq \frac{e^2}{4\pp}.
\end{equation}
One sees that $E_\text{c}$ is much greater than the Schwinger limit and
breaks the weak field condition in eq.\ \eqref{eq:weakcondition}, as have
been noted in e.g.~\cite{kawai2015}. Therefore, the obtained results are
not valid when the field goes close to the critical
limit, and cannot answer the question whether a vacuum cascade happens
near the critical field strength.

\subsection{Single-pulse field $E(t)=E \sech^2(\omega t)$}%


For a generic 1D dynamic field, we assume that the Wilson loop is of the following form
\begin{equation}
\braket{W(\gamma)}=\sfun{\exp}{\frac{e^2}{4}\lambda(\gamma)},
\quad
\gamma=\frac{m\omega}{e E},
\end{equation}
where $\lambda(\gamma)$ is an \emph{enhancement factor} 
with respect to the case of a constant external field
in eq.\ \eqref{eq:wlconstant}.
This name of $\lambda(\gamma)$ is seen to be appropriate from the fact that $\lambda(\gamma)$ is a monotonically decreasing function and tends to unit at adiabatic limit $\gamma\to 0$.

\begin{figure} 
\begin{center}
\begin{subfigure}{\linewidth}
\begin{center}
\includegraphics{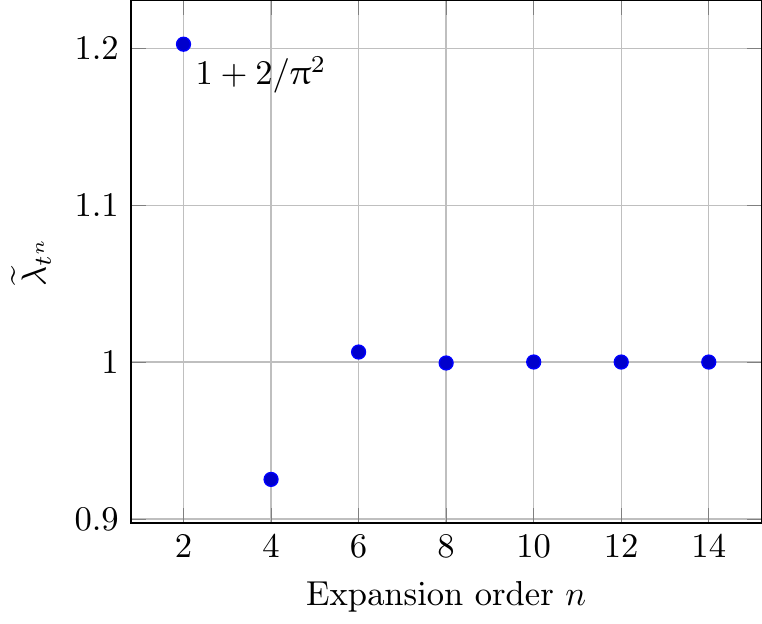}
\end{center}
\end{subfigure}

\begin{subfigure}{\linewidth}
\begin{center}
\includegraphics{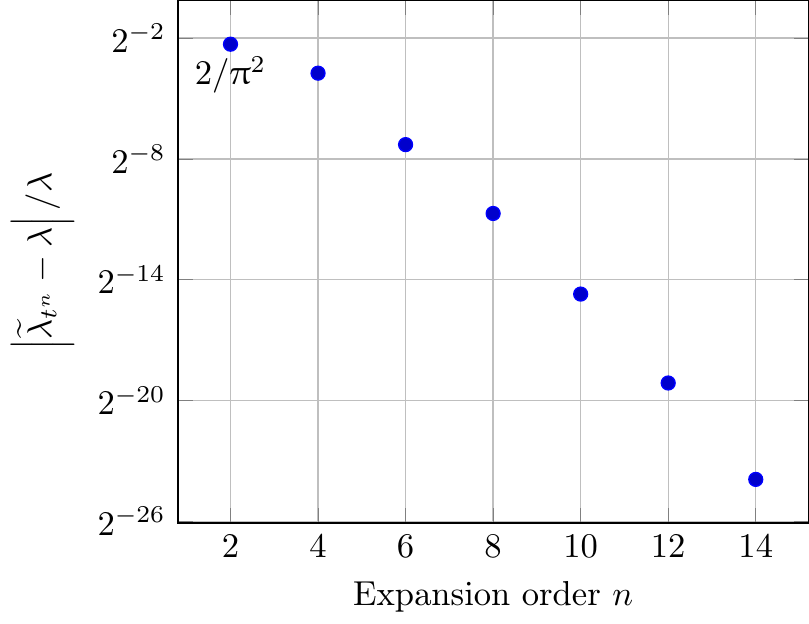}
\end{center}
\end{subfigure}
\end{center}
\caption{
First seven orders of $\widetilde{\lambda}_{t^n}$ and their deviation from $\lambda$ for a constant electric field.
Because of symmetry of the integrals, all the odd orders vanish.
\label{fig:texpansion}}
\end{figure}

Before applying nonlinear approximation schemes as we discussed in sec.\ \ref{ssec:est-A} to single-pulse field, we apply it to the 
case of a constant field. The integrand after $t$-expansion becomes
\begin{equation*}
    -\frac{R^2 \left(t^2-2\right)}{2 \left(R^2 t^2+\varepsilon ^2\right)}
\end{equation*}
while the subtracting term eq.\ \eqref{eq:substract} will not change.
Then repeating the similar procedure, one gets
\begin{equation*}
    \widetilde{\lambda}=1+\frac{2}{\pp^2}
\end{equation*}
which leads to a $2/\pp^2$ deviation compared to
eq.\ \eqref{eq:wlconstant}.
This term comes from $-4$ in ${}^2\!\mathcal{A}^{(0)}$ and does not depend on 
the specific form of worldline path.  
Later we will see that this $2/\pp^2$ deviation happens in both cases considered
in 2nd-order $t$-expansion, which
is caused by accuracy of estimation method, 
thus it approaches to zero as the approximation order increases, see fig.\ \ref{fig:texpansion}.
In other words, the emergence of finite terms are expected for each orders, and 
the higher-order contribution
should cancel $2/\pp^2$ from the 2nd-order. 
Based on this consideration we remove $2/\pp^2$ 
derivation directly in the final results,
which should not be confused with the counter term eq.\ \eqref{eq:regulator}.

Let us see the first nontrivial example, pair production in the
single-pulse field \cite{popov1971,dunne2005b}.
The instanton has been evaluated explicitly to be
\begin{equation}\label{eq:singlepulse}
\begin{split}
x_0  &= \frac{1}{\omega }
	\sfun{\arcsin}{\frac{\gamma  }{\sqrt{1 + \gamma^2}}\sin (u)},
\\
x_1  &= \frac{1}{\omega }
	\frac{1}{\sqrt{1 + \gamma^2}}\,
	\sfun{\arcsinh}{\gamma  \cos (u)},
\end{split}
\end{equation}
and the worldline action for this path reads
\begin{equation}
S_0 = \frac{\pp m^2}{e E}\frac{2}{1+\sqrt{1+\gamma^2}}.
\end{equation}
The weak-field condition leads to
\begin{equation}
\label{eq:weakfield}
E\ll \frac{2\pp E_\text{s}}{\sqrt{1+\gamma^2}}.
\end{equation}
At the nonperturbative region \cite{gelis2016} $\gamma\ll 1$,
the weak-field condition reduces to $E\ll 2\pp E_\text{s}$. In other words, the dynamics of field also decreases the upper limit of weak-field condition. 

\begin{figure}
\begin{center}
\begin{subfigure}{\linewidth}
\begin{center}
\includegraphics{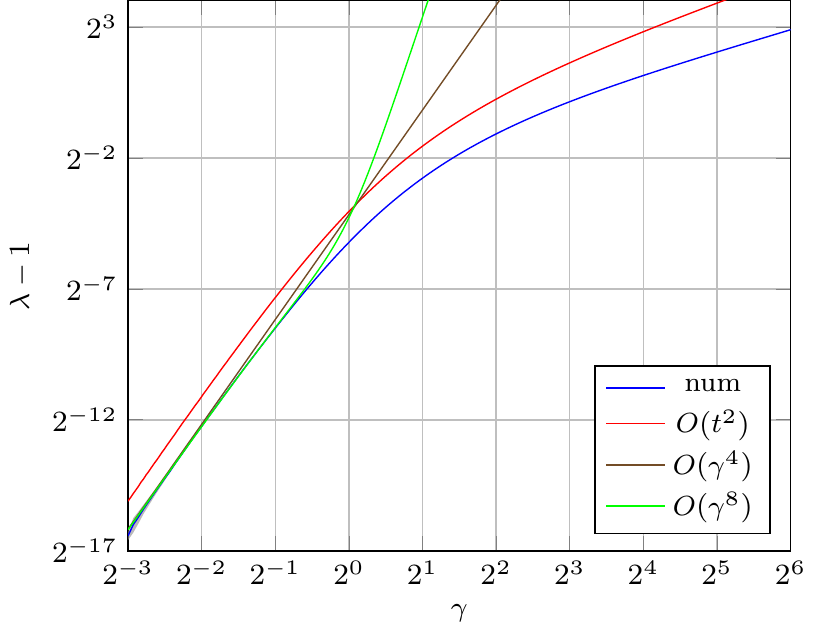}
\end{center}
\subcaption{$\gamma \le 2^{6}$, double logarithm axes
}
\end{subfigure}

\begin{subfigure}{\linewidth}
\begin{center}
\includegraphics{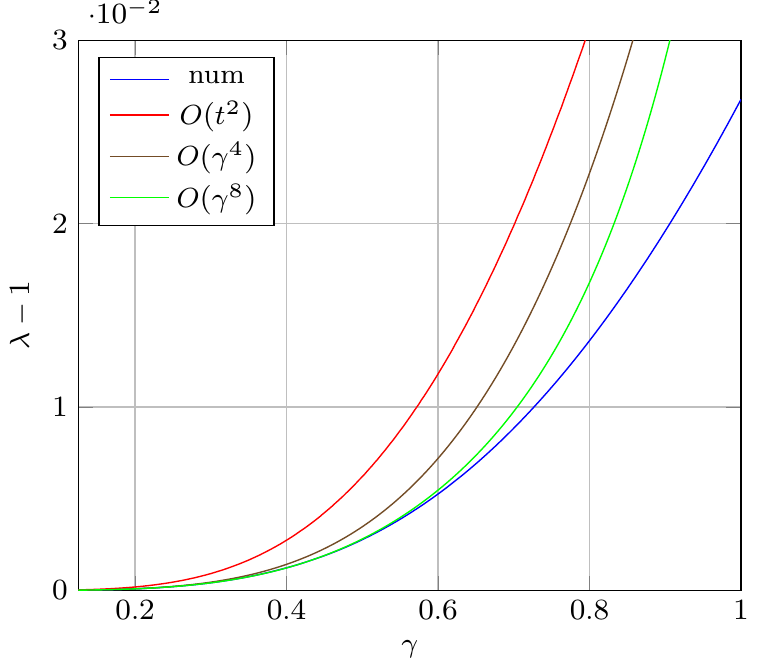}
\end{center}
\subcaption{$\gamma \le 1$, double linear axes
}
\end{subfigure}
\end{center}
\caption{
The dependence of enhancement factor $\lambda$ on $\gamma$ in a single-pulse field,
showing a comparison of numerical result with estimation of
$t$-expansion as well as $\gamma$-expansions.
The shadow indicates the numerical error in extrapolating
$\varepsilon\to0^+$ with $95\%$ confidence.
\label{fig:singlepulse0}}
\end{figure}

Now we turn to the the Wilson loop. 
The corresponding perimeter law has a closed form
\begin{equation}
\label{eq:eta}
\dva\mathcal{A}=\frac{2 \pp ^2 \eta }{ \omega  \e  },
\qquad 
\eta=\frac{ \gamma  }{ \sqrt{1+\gamma^2} }.
\end{equation}
With the $t$-expansion up to $\rfun{O}{t^2}$, eq.\ \eqref{eq:wilsonphase}
reads
\begin{equation}\begin{split}
{}^2\!\mathcal{A} &\approx \frac{\pp }{\omega  \e\gamma  \sqrt{1 + \gamma^2}  }
 \Bigg\{
 -\pp \omega  \e \gamma  \rbr{2 + \gamma^2}+
 \\
&
+\Big[\gamma^2 \rbr{\omega ^2 \varepsilon ^2\sqrt{1 + \gamma^2} +4}
 +2 \omega ^2 \e ^2\sqrt{1 + \gamma^2} \Big]\times\\
&\qquad\qquad\times \rfun{\arctan}{\frac{\pp  \gamma }{\omega  \e\sqrt{1 + \gamma^2}  }}  \Bigg\}.
\end{split}\end{equation}
which leads to
\begin{equation*}
\widetilde{\lambda}(\gamma)=\frac{1}{2} 
\rbr{\frac{2 + \gamma^2}{\sqrt{1 + \gamma^2}}+\frac{4}{\pp ^2}}.
\end{equation*}
One may note that $\widetilde{\lambda}\to 1+2/\pp^2$ as $\gamma$ approaches zero. The $2/\pp^2$ derivation is predicted and has to be subtracted, i.e.\ the enhancement factor is then in the weak-field condition 
\begin{equation}
\label{eq:enhfactor1}
\lambda(\gamma)=\frac{1}{2} 
\rbr{\frac{2 + \gamma^2}{\sqrt{1 + \gamma^2}}}.
\end{equation}
This operation guarantees the condition: $\lambda\to 1$ as $\gamma \to 0$.
Since $\lambda(0)>1$, the factor amplifies the contribution from the
Wilson loop in eq.\ \eqref{eq:wlconstant}, so that the pair-production
rate is no longer exponentially suppressed, see fig.\ \ref{fig:singlepulse0}.

Alternatively, the $\gamma$-expansion can be worked out as a power series
of $\gamma$, where eq.\ \eqref{eq:singlepulse} has been used by replacing
$\omega = \gamma/R$.
After removing the poles at each order of $\lambda$, one obtains
\begin{equation}
\label{eq:enhfactor1-gamma}
\lambda(\gamma)=1+\frac{1}{18} \gamma^4 - \frac{1}{18} \gamma^6 + 
\frac{443}{8640} \gamma^8 + \rfun{O}{\gamma^{10}}.
\end{equation}

The exponential factor in the production rate is evaluated as
\begin{equation}
    \Gamma \sim
    \rfun{\exp}{-
    \frac{\pp E_\text{s}}{E}\frac{2}{1+\sqrt{1+\gamma^2}}+\frac{\pp \alpha}{2}
    \frac{2 + \gamma^2}{\sqrt{1 + \gamma^2}}}
\label{eq:rate1}
\end{equation}
The first part is the main contribution from instanton action, the second part arises due to the Wilson loop correction from $t$-expansion.
If the critical field is defined as saddle point, at which the exponential suppression is precisely zero,
then one could have
\begin{equation}
E_\text{c}= \frac{E_\text{s}}{\a}  
\frac{4 \sqrt{\gamma ^2+1}}{\left(\gamma ^2+2\right) \left(\sqrt{\gamma ^2+1}+1\right)}
< 137 E_\text{s}.    
\label{eq:small-gamma}
\end{equation}
On the one hand,
as in the case with constant field, eq.\ \eqref{eq:small-gamma} also breaks
the weak-field condition at nonperturbative region $\gamma\ll1$. 
On the other hand,
in contrast with the case in constant field, the exponential factor of Wilson loop is no longer
constant, such that for given $E$ its contribution for production rate becomes dominant as $\gamma$ increases. 
In addition, both estimation eq.\ \eqref{eq:rate1} (or eq.\ \eqref{eq:enhfactor1}) and numerical result seem to be divergent
as $\gamma$ approaches $+\infty$,
even if the pre-exponential factor of Feynman integral were taken into account \cite{dunne2006}. It implies that Wilson loop ought to be of a pole at $\gamma\to+\infty$, and loses its meaning at this point, where the instanton trajectory collapses to a singular point.

\subsection{Sinusoidal field $E(t)=E \cos(\omega t)$}

\begin{figure}
\begin{center}
\begin{subfigure}{\linewidth}
\begin{center}
\includegraphics{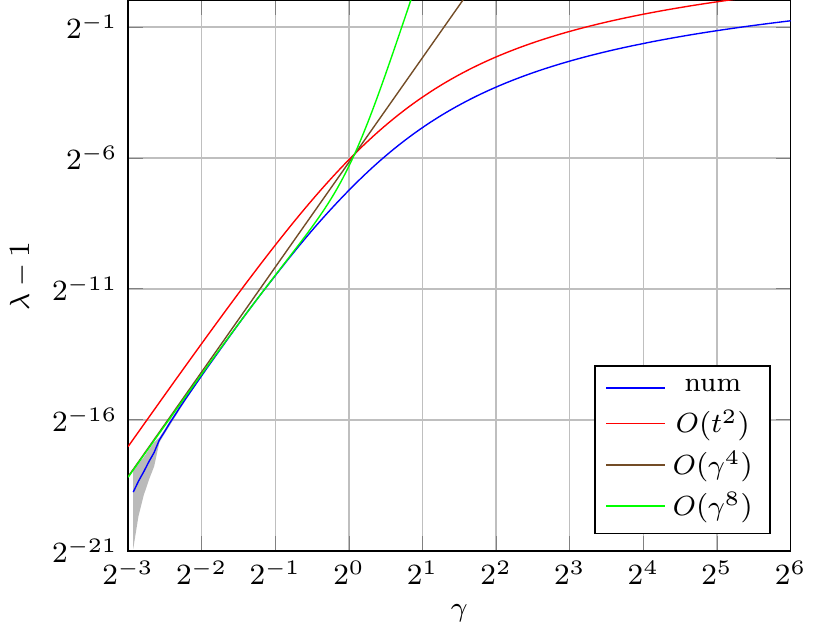}
\end{center}
\subcaption{$\gamma \le 2^{6}$, double logarithm axes
}
\end{subfigure}

\begin{subfigure}{\linewidth}
\begin{center}
\includegraphics{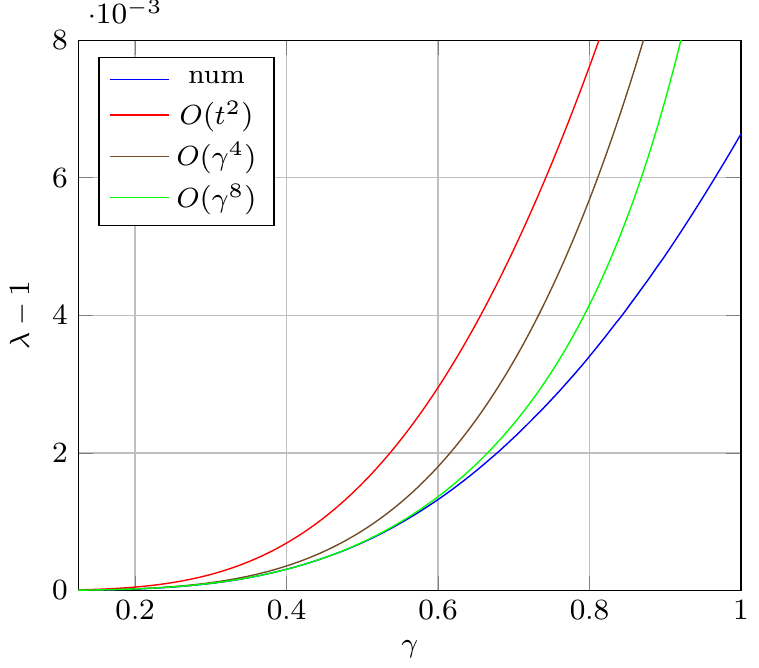}
\end{center}
\subcaption{$\gamma \le 1$, double linear axes
}
\end{subfigure}
\end{center}
\caption{The dependence of enhancement factor $\lambda$ on $\gamma$ in a 
sinusoidal field, showing a comparison of numerical result with estimations of 
$t$-expansion and fourth-order $\gamma$-expansions.
The shadow indicates the numerical error in extrapolating
$\varepsilon\to0^+$ with $95\%$ confidence.
\label{fig:sinusoidal0}}
\end{figure}

The second example is a sinusoidal field
\cite{brezin1970,popov1971,dunne2005b}.
The coordinates of the instanton can be represented by special functions as
\begin{equation}\begin{split}
x_0 &=\frac{1}{\omega }
\sfun{\arcsinh}{\eta\, \rfun{\sd}{\frac{2 }{\pp }\rfun{K}{\eta^2} u\bigg |\eta^2}},\\
x_1 &=
\frac{1}{\omega }
\sfun{\arcsin}{\eta\,
	\rfun{\cd}{\frac{2 }{\pp }\rfun{K}{\eta^2} u\bigg |\eta^2}},
\end{split}\end{equation}
where $\rfun{K}{\cdot}$ is the complete elliptic integral,
${\sd}(\cdot)$ and ${\cd}(\cdot)$ are Jacobi elliptic functions. 
The instanton action is given by
\begin{equation}
\label{eq:insacsin}
S_0 = \frac{m^2}{e E} \frac{4\sqrt{1+\gamma^2}}{\gamma^2}
\sbr{\rfun{K}{\eta^2}-\rfun{E}{\eta^2}}.
\end{equation}
Repeating the procedure above, one obtains the corresponding perimeter law 
in a closed form
\begin{equation}
\dva\mathcal{A}=
\frac{4 \pp  \eta  }{\omega  \e  }
\rfun{K}{\eta^2}.
\end{equation}
The exponent in the $t$-expansion reads
\begin{equation}
\begin{split}
\mathcal{A}\approx
\frac{8 \eta  }{\omega  \e   } &\rfun{K}{\eta^2} 
	\sfun{\arctan}{\frac{2 \eta }{ \omega  \e}
	 \rfun{K}{\eta^2}}+
\\
&
+\sfun{E}{\rfun{\am}{2 \rfun{K}{\eta^2}|\eta^2}|\eta^2}\times\\
&\times
 \cbr{\frac{2\omega  \e  }{  \eta}
 \sfun{\arctan}{\frac{2 \eta }{\omega  \e } \rfun{K}{\eta^2}}
 -4 \rfun{K}{\eta^2}},
\end{split}
\end{equation}
in which $\rfun{\am}{\cdot}$ is the Jacobi amplitude function. 
Thus at the week-field approximation $E\ll 4 E_{\text{s}}\rfun{K}{\eta^2}/\sqrt{1+\gamma^2}$ \cite{dunne2005b}, one has
\begin{equation}
\lambda(\gamma)=\frac{2 }{\pp ^2}
	\rfun{K}{\eta^2} \sfun{E}{\rfun{\am}{2 \rfun{K}{\eta^2}|\eta^2}|\eta^2}.
\label{eq:enhfactor2}
\end{equation}
Where the deviation $2/\pp^2$ has been subtracted.
In this case, $\lambda(\gamma)$ is also greater than unity and a non-trivial
function depending on external field, and it tends to $1$ as $\gamma$ approaches zero.
Alternativeszaly, the $\gamma$-expansion is implemented by expanding in $\eta$
first. Removing the divergences in $\varepsilon$, one obtains
\begin{equation}
    \lambda= 1 + \frac{\gamma^4}{72}  - \frac{\gamma^6}{72} +
    \frac{217 \gamma^8}{17280} + \rfun{O}{\gamma^{10}}.
\end{equation}
The results are shown in fig.\ \ref{fig:sinusoidal0}. 

In addition, similar to the example in the last subsection $\lambda$ is also divergent as $\gamma \to \infty$, even if the
pre-exponential factor is considered \cite{dunne2006}.
The exponent factor in the production rate is then given by
\begin{equation}
\label{eq:rate2}
\begin{split}
\Gamma  \sim \exp\!\bigg\{&
- \frac{E_s}{E} \frac{4\sqrt{1+\gamma^2}}{\gamma^2}
\sbr{\rfun{K}{\eta^2}-\rfun{E}{\eta^2}}+\\
&+
\frac{2\alpha}{\pp }
	\rfun{K}{\eta^2}\rfun{E}{\rfun{\am}{2 \rfun{K}{\eta^2}|\eta^2}|\eta^2}
\bigg\},
\end{split}\end{equation}
For small $\gamma$, one gets
\begin{equation}
E_\text{c} \sim \frac{E_\text{s}}{\a} 
\left(1-\frac{\gamma ^2}{8}\right) +O(\gamma^4)<137 E_s,
\end{equation}
which breaks the weak-field condition at nonperturbative region as well.

From above three examples,
one may note that, first, 
the weak-field condition in the non-perturbative ranges $\gamma<1$ is inevitably broken at strong coupling, which makes the vacuum cascade around the critical field ambiguous \cite{kawai2015}; and
second, 
the correction due to the Wilson loop in dynamic fields
is a monotonically increasing function with respect to $\gamma$ and
diverges as $\gamma\to\infty$.

\section{Holographic Schwinger effect with dynamic field}
\label{sec:duality}

In order to answer the question, if the vacuum cascade for strong coupling happens
as the strength of time-dependent field goes close to the critical limit \cite{kawai2015}, we
consider a similar effect in the context of gauge/gravity duality, where the gauge
field theory refers to an $\mathcal{N}=4$ $\mathsf{SU}(N+1)$ supersymmetric
Yang--Mills theory on the 4D boundary of an $\mathrm{AdS}_5\times S^5$ space,
and the quantum gravity is a type \rom{2}B superstring theory in 
the bulk of the $\mathrm{AdS}_5\times S^5$.
The same as the case with constant field, we expect that the string theory could shed some light on the catastrophic vacuum cascade through the duality principle.

According to the Semenoff and Zarembo's holographic setup
\cite{semenoff2011,sato2013a,sato2013b,sato2013c,sato2013d,kawai2014},
the exponential factor in the production rate of the gauge field is obtained
from the superstring counterpart by the area of the string worldsheet
attached to a probe D3--brane, i.e.\
\begin{equation}
    \Gamma\sim \rfun{\exp}{-S_\text{NG}-S_{B_2}},
\end{equation}
where $S_\text{NG}$ is the Nambu--Got\=o (NG) action
\cite{Nambu1970a,Goto1971}
\begin{equation}
\label{eq:nambu-goto}
    S_\text{NG}= T_F \int \dif{}^2 \sigma \sqrt{\vbr{\det G_{\alpha\beta}}}
\end{equation}
depending on the induced metric
\begin{equation}
    G_{\alpha\beta} =
    g_{MN}\frac{\partial x^M}{\partial\sigma^\alpha}
    \frac{\partial x^N}{\partial\sigma^\beta}
    ,
\end{equation}
and $S_{B_2}$ is the Kalb--Ramond
\cite{Kalb1974}
2-form
(or NS--NS, where NS is the abbreviation of Neveu--Schwarz
\cite{Neveu1971}
)
as an string interaction term,
\begin{equation}
    S_{B_2} = -T_F \int \dif{}^2 \sigma\,
    B_{MN} \frac{\partial x^M}{\partial\sigma}
    \frac{\partial x^N}{\partial\tau}.
\label{eq:ns-ns-def}
\end{equation}
In eqs.\ \eqref{eq:nambu-goto} and \eqref{eq:ns-ns-def},
$\sigma^\alpha=(\tau, \sigma)$ are the coordinates on the
string worldsheet, $x^M=(x_\nu,r,\phi_a)$ are the coordinates
of the $10$-D $\mathrm{AdS}_5\times S^5$ space with metric $g_{MN}$,
and $G_{\alpha\beta}$ is the induced metric.

In the Semenoff--Zarembo construction, the worldsheet ends 
on the probe D3--brane with a boundary, taking the same shape 
as the worldline instanton. Hence the essential problem is 
converted to compute the on-shell action of string in Euclidean 
$\mathrm{AdS}_5$ with the given boundary. Note that the Nambu--Got\=o
action is proportional to the worldsheet area, and extremising
the area leads to a minimal surface. In other words, calculation
of the exponential factor now corresponds to a Plateau's problem
in the framework of gauge/gravity duality.

\subsection{Constant electric field}
\label{ssec:duality-con}

The worldline instanton in a constant field is a circle, thus the worldsheet can be parametrised by
\begin{equation}
\label{eq:duality-con-10}
    x_0 = \sigma \cos(\tau),\qquad
    x_1 = \sigma \sin(\tau),
\end{equation}
which is different than the choice in \cite{kawai2015}, 
and the two parameterisations have different chirality,
i.e.\ $\mathrm{sgn}(J)=-1$, where $J$ is the Jacobian.
Therefore, the orientation of the Kalb--Ramond coupling is also reversed.
The Nambu--Got\=o action in our parameterisation becomes 
\begin{equation}
    S_\text{NG}= T_F \int_0^{2\pp} \dif \tau\,
    \int_{0}^{R} \dif \sigma\,
    \sqrt{\vbr{\det G_{\alpha\beta}}},
\end{equation}
where $R \equiv \sigma(r_0)$, $r_0$ is the location of the probe D3--brane.
$r(\sigma)$ can be analytically as \cite{xiao2008}
\begin{equation}
   r = \frac{L^2}{\sqrt{\frac{L^4}{r_0^2}+f(\sigma)}},
\label{eq:duality-con-20}
\end{equation}
where $f(\sigma)=R^2-\sigma^2=0$ is the worldline
instanton on the D3--brane. Substituting it into the Nambu--Got\=o action, one obtains
\begin{equation}
\label{eq:circleNG}
\begin{split}
   S_\text{NG} &=2\pp T_F L^2 \sqrt{\frac{L^4}{r_0^2}+R^2}
   \int^{R}_0 
   \frac{ \sigma \,\dif \sigma}
   {\left(\frac{L^4}{r_0^2}+R^2-\sigma ^2\right)^{3/2}}\\
   &=
   2 \pp  L^2 T_F \left(\sqrt{\frac{r_0^2 R^2}{L^2}+1}-1\right).
   \end{split}
\end{equation}
Furthermore, the NS--NS term reads
\begin{equation}
    S_{B_2}= \rfun{\mathrm{sgn}}{J}\int ^{2\pp}_0
    \dif \tau \int^{R}_0\dif\sigma\, E_0 \sigma
    =-\pp E_0 R^2,
\end{equation}
where $R$ can be fixed by extremising the total action, yielding
\begin{equation}
    R=\frac{L^2}{r_0}\sqrt{\frac{E_\text{c}^2}{E_0^2}-1},\qquad
    E_\text{c}=T_F\frac{r_0^2 }{L^2}.
\end{equation}
Where $E_\text{c}$ is defined as critical values.
The exponential factor in the production rate can now be solved as
\begin{equation}
\label{eq:decayrate0}
    \Gamma\sim \sfun{\exp}{
    -\frac{\sqrt{\mathfrak{g}}}{2} \rbr{
     \sqrt{\frac{E_\text{c}}{E_0}}
     -\sqrt{\frac{E_0}{E_\text{c}}} }^2},
\end{equation}
in which the string and spacetime parameters have been replaced by
the ones of the gauge field via $T_F=\sqrt{\mathfrak{g}}/(2\pp L^2)$. This is the result obtained in \cite{semenoff2011} and agrees with the Schwinger's formula in the weak-field  limit.

\subsection{Estimation of Single-pulse field $E(t)=E \sech^2(\omega t)$}
\label{ssec:duality-pul}

\begin{figure} 
\begin{center}
\includegraphics{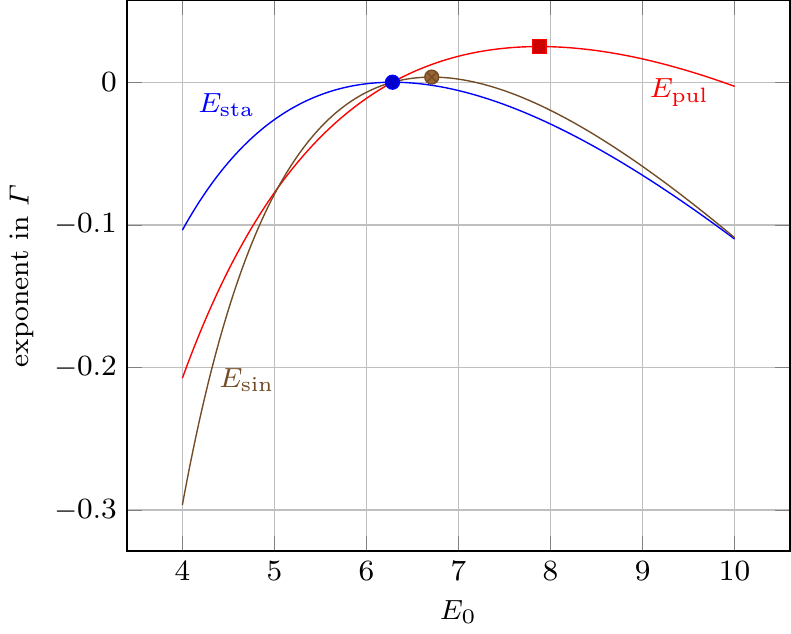}
\end{center}
\caption{The exponents of the vacuum decay rate, where the critical 
field is defined by the maxima, which are emphasised by dots. For constant 
field, the exponent is given by eq.\ \eqref{eq:decayrate0} and plotted in
blue.
For single-pulse field in eq.\ \eqref{eq:decayrate1}, the colour is 
red.
For sinusoidal field shown in eq.\ \eqref{eq:decayrate2}, it is plotted in 
brown.
\label{fig:decarate}}
\end{figure}

The instanton path as the worldsheet boundary on the D3--brane in a
single-pulse field has been shown in eq.\ \eqref{eq:singlepulse}. Thus one can
parametrise the worldsheet by using $\sigma$ and $u$, i.e.\
\begin{equation}
\begin{split}
    x_0 &=  \frac{\sigma}{\gamma }\,
    \sfun{\arcsin}{\frac{\gamma}{ \sqrt{1+\gamma^2}} \sin (u)},\\
    x_1 &=  \frac{\sigma}{\gamma \sqrt{1+\gamma^2}}\,
    \sfun{\arcsinh}{\gamma \cos (u)},
\end{split}
    \label{eq:duality-pul-10}
\end{equation}
where $\gamma$ is regarded as initial information and not relevant to the scale $r$.
The simplicity of Semenoff--Zarembo construction for constant field led us to speculate that the similar production rates could have been obtained by repeating above procedure. 
However, it is not anything like worldline instanton, 
the string worldsheets are not exactly integrable for dynamic fields in our cases. Thus to make an effect estimation, 
we expand the instanton 
at adiabatic limit, i.e. $\gamma\to 0$,
\begin{equation}
\label{eq:boundary1}
\begin{split}
    x_0&\sim \sigma  \sin (u)-\sigma\frac{\gamma^2 }{6}   \cos (u) \left[\cos ^2(u)+3\right]+\rfun{O}{\gamma ^4},\\
    x_1&\sim \sigma  \cos (u)-\sigma\frac{\gamma^2 }{24}   [9 \sin (u)+\sin (3 u)]+\rfun{O}{\gamma ^4},
\end{split}
\end{equation}
namely the zeroth order of $\gamma$ is just the circle boundary eq.\ \eqref{eq:duality-con-10}.
Therefore, we will treat the complete worldline boundary as perturbation
around the circle. The $r$ component can be estimated by noting that
\begin{equation}
r=\frac{L^2  }{\sqrt{ \frac{L^4}{r_0^2} + f(x_0,x_1)}},
\end{equation}
where $f(x_0,x_1)=0$ is the instanton path.
In a single-pulse field, 
it reads
\begin{equation}
\begin{split}
   f(x_0,x_1)=
   \frac{\sigma _0^2 }{2 \gamma^2}
   \Bigg[\gamma^2+&\left(1 + \gamma^2\right) \rfun{\cos}{\frac{2 \gamma  }{\sigma _0}x_0}  -\\
   &-
   \rfun{\cosh}{\frac{2 \gamma  \sqrt{1 + \gamma^2} }{\sigma_0}x_1}\Bigg].
\end{split}
\end{equation}
In the limit $\gamma\to 0$, $f(x_0,x_1)$ reduces to a circle 
\begin{equation}
\begin{split}
   f(x_0,x_1)\sim \sigma _0^2-x_0^2 &-x_1^2
   +\frac{\gamma^2 }{3 \sigma _0^2}\left(x_0^2+x_1^2\right)\times\\
   &\times\left(-3 \sigma _0^2+x_0^2-x_1^2\right) +
   \rfun{O}{\gamma ^4}.
\end{split}
\end{equation}
Hence the Nambu--Got\=o action is formulated as
\begin{equation}
    S_{\text{NG}}=S^0_{\text{NG}}
    -\pp  T_F\frac{ \gamma^2 r_0 \sigma _0^2 }{\sqrt{\frac{L^4}{r_0^2}+\sigma _0^2}},
\end{equation}
where $S^0_{\text{NG}}$ is given in eq.\ \eqref{eq:circleNG}.
On the other hand, the NS--NS part right now becomes
 \begin{equation}
    S_{B_2}= S_{B_2}^0
    +\frac{3}{2}\pp  E_0  \gamma^2 \sigma_0^2.
\end{equation}
The $\sigma_0$ is fixed as the stationary point of the total action, 
\begin{equation}
    \sigma_0\sim
    \frac{L^2 }{r_0}
    \sqrt{\frac{E_\text{c}^2}{E_0^2}-1}+\gamma^2
  \frac{ L^2 }{2 r_0 }
  \frac{\frac{2 E_\text{c}^2}{E_0^2}-1}{\sqrt{\frac{E_\text{c}^2}{E_0^2}-1}}
  +O\left(\gamma ^3\right).
\end{equation}
The exponential factor in the production rate with correction up to the second order of $\gamma$ is 
\begin{equation}
\label{eq:decayrate1}
\begin{split}
    \Gamma\sim 
   \exp\Bigg[-
    \frac{\sqrt{\mathfrak{g} } }{2 }&
    \left(
    \sqrt{\frac{E_0}{E_\text{c}}}
    -\sqrt{\frac{E_\text{c}}{E_0}}
    \right)^2-\\
    &-\gamma^2\frac{ \sqrt{\mathfrak{g} } }{4 }
    \left(\frac{E_\text{c}}{E_0}-\frac{E_0}{E_\text{c}}\right)
    +\rfun{O}{\gamma ^4}\Bigg].
\end{split}
\end{equation}
here $\mathfrak{g}=g^2_{\text{YM}}N$ denotes the effective coupling at large 't Hooft limit. In eq.\ \eqref{eq:decayrate1}, the first part in the exponent comes from the circular
boundary, while the second term arises from the deformation, see fig.\
\ref{fig:decarate}.

If one defines the critical field as stationary point of exponent, the correction (up to $O(\gamma^2)$) due to the time-dependence of the background field leads a lager critical value comparing with constant field, which is fixed by
\begin{equation}
    E_0\to E_{\text{c}}\sqrt{
    \frac{2+\gamma^2}{2-\gamma^2}
    } ,
\end{equation}
beyond which the decay rate decreases as the similar as eq.\ \eqref{eq:decayrate0}.
The enhancement up to $O(\gamma^2)$ can also be noted from the negative sign of $\gamma^2$ in eq.\ \eqref{eq:decayrate1}. In other words, the correction plays a role of suppression of pair production. 

\subsection{Estimation of sinusoidal field $E(t)=E \cos(\omega t)$}


The worldline instanton in this case can be parametrised as
\begin{equation}\begin{split}
x_0 &=\frac{\sigma}{\gamma }\,
\sfun{\arcsinh}{\eta\, \rfun{\sd}{\frac{2 }{\pp }\rfun{K}{\eta^2} u\bigg |\eta^2}},
\\
x_1 &=
\frac{\sigma}{\gamma }\,
\sfun{\arcsin}{\eta\,
	\rfun{\cd}{\frac{2 }{\pp }\rfun{K}{\eta^2} u\bigg |\eta^2}},
\end{split}\end{equation}
where $\eta$ has been defined in eq.\ \eqref{eq:eta}.
The Maclaurin series of Jacobian elliptic functions in $\gamma$ gives
\begin{equation}\begin{split}
    x_0   \sim
    \sigma  \sin (u)-&
    \sigma\frac{\gamma^2}{48}    [9 \sin (u)+\\
    &+\sin (3 u)+12 u \cos (u)]+\rfun{O}{\gamma ^4},\\
    x_1   \sim \sigma  \cos (u)
    - &\sigma\frac{\gamma^2}{48}   [-12 u \sin (u)+\\
    &+15 \cos (u)+
    \cos (3 u)]
    +\rfun{O}{\gamma ^4}.
\end{split}
\end{equation}
The instanton path on the D3--brane in this case is
\begin{equation}\begin{split}
    f(x_0,x_1)=
   & \frac{\sigma _0^2 }{2 \gamma^2 \left(1 + \gamma^2\right)}
    \bigg[\gamma ^4+\left(1 + \gamma^2\right)^2 \rfun{\cos}{\frac{2 \gamma  }{\sigma _0}x_0}+\\
    &+\left(21 + \gamma^2\right)\rfun{\cosh}{\frac{2 \gamma  }{\sigma _0}x_1}-2(1+ \gamma^2)\bigg]
\end{split}
\end{equation}
and its expansion for small $\gamma$ reads
\begin{equation}\begin{split}
   f(x_0,x_1)\sim &\sigma _0^2-x_0^2-x_1^2+\gamma^2\times\\
   &\times
   \left(\frac{x_0^4-x_1^4}{3 \sigma _0^2}-x_0^2-x_1^2\right)+\rfun{O}{\gamma ^4}.
\end{split}
\end{equation}
The second order terms for the Nambu--Got\=o action 
\begin{equation}\begin{split}
      S_{\text{NG}}^{( 2)}  =&\frac{\pp   T_F \gamma^2}{2 L^4 \sqrt{L^4+r_0^2 \sigma _0^2}}
      (r_0^4 \sigma _0^4-L^8-\\
      &-2 L^4 r_0^2 \sigma _0^2+L^6 \sqrt{L^4+r_0^2 \sigma _0^2}),  
\end{split}
\end{equation}
and the NS--NS term are
\begin{equation}
      S_{B_2}^{( 2)} =\pp  \gamma^2 E_0 \sigma_0^2.
\end{equation}
Up to the second order of $\gamma$, $\sigma_0$ is given by
\begin{equation}
   \sigma_0=\frac{L^2}{r_0}
   \sqrt{\frac{E_\text{c}^2}{E_0^2}-1}
   +\gamma^2\frac{ L^2 }{2 r_0 }
   \frac{\frac{3 E_\text{c}^4}{2E_0^4}-1}{\sqrt{\frac{E_\text{c}^2}{E_0^2}-1}}
   +O(\gamma^4).
\end{equation}
Finally, the exponential factor in the production rate is given by
\begin{equation}\begin{split}
    \Gamma\sim 
    \exp\Bigg[&-
    \frac{\sqrt{\mathfrak{g} } }{2 }
    \left(
    \sqrt{\frac{E_0}{E_\text{c}}}
    -\sqrt{\frac{E_\text{c}}{E_0}}
    \right)^2-\\
    &-\gamma^2\frac{ \sqrt{\mathfrak{g} } }{4 }
    \left(
    \frac{E_\text{c}^3}{E_0^3}
    -2\frac{E_\text{c}}{E_0}+1\right)
    +\rfun{O}{\gamma ^4}\Bigg],
\label{eq:decayrate2}
\end{split}
\end{equation}
see fig.\ \ref{fig:decarate}.
The critical field should be greater than $E_\text{c} = T_F r_0^2/L^2$, i.e.
\begin{equation}
    E_0\to 
    \frac{\sqrt{1-\gamma^2+\sqrt{\gamma ^4+4 1 + \gamma^2}} }{\sqrt{2}}E_{\text{c}}
\end{equation}
and one sees that dynamics of the external field also enhances the critical field as before.

\section{Conclusion and Discussion}


In this paper, the scalar Schwinger effect for dynamic fields at strong coupling and weak-field limit has been studied,
by first using the field-theoretical method of worldline instantons. 
A non-trivial contribution 
to the production rate is discovered by evaluating the Wilson loop along the instanton path, which depends on the Keldysh adiabaticity parameter $\gamma$.
Thus one may expect that such correction may save the weak-field condition in strong coupling.  
However after computations, we find that the introduction of the correction term also
leads to a contradiction to the weak-field condition
near the critical field strength.
 
We note also that the correction from Wilson loop is a monotonically increasing function
with respect to $\gamma$, which makes the contribution for production rate from Wilson loop become dominant as $\gamma$ increases.
Moreover both $t$-expansion and numerical calculation suggest a divergent value as $\gamma$ approaches infinity, 
even if the pre-exponential factor of Feynman integral is considered. 
One possible explanation is that the Wilson loop loses its meaning at $\gamma\to \infty$, because the instanton trajectory collapses to a singular point. 

In order to clarify the vacuum cascade beyond the weak-field condition, 
in the context of an $\mathcal{N}=4$ supersymmetric Yang--Mills
theory, the production rate is calculated by the gauge/gravity duality,
according to which the instanton action has a string counterpart of the classical string action
in Euclidean $\mathrm{AdS}_3$, where the boundary
on the probe D3--brane is given by the instanton path. Thus the 
problem is converted to solving the classical motion of string with 
Dirichlet boundary conditions. 
However the string worldsheets for dynamic fields are not 
integrable as in the worldline instantons. 
To provide an 
explicit estimation, we treat the specific worldsheets as perturbations 
around the one with circle boundary, which had been solved exactly. 
Such an expansion is an adiabatic approximation, it is
practical and realistic, because only low-frequency laser (comparing with electron mass) is currently operational.
The obtained decay rates in the two examples with dynamic fields are
similar concave functions as in the case with constant field,
but the critical fields increase considerably.
In other words, up to $O(\gamma^2)$ the correction due to the dynamics of electric field suppresses the pair production, which is opposite to cases of worldline instantons.

\appendix





\section{The validity of $t$-expansion}
\label{app:texpansion}

For an approximation theory $\{g_n(x)\}$ of full function $f(x)$, we say it is valid if $\lim_{n\to \infty}g_n=f$ in some domain. 
The purpose of $t$-expansion is to provide a nonlinear approximation for $A_\e(\gamma)$ by establishing a  sequence of functions $\{{}^n\!A_\e(\gamma)\}$, which satisfies 
$\lim_{n\to\infty} {}^n\!A_\e =A_\e $.

In order to establish nonlinear approximation, we note that the integrand of $A_\e$ can be expanded as
\begin{equation}\label{eq:integrand}
\begin{split}
    T(t)&= \frac{ x'(s) \cdot x'(s+t) }{[x(s+t)-x(s)]^2+\e^2}\\
      &=
        \frac{\sum^\infty_{k=0} a_k t^k}{\e^2+\sum^\infty_{k=2}c_k t^k},
\end{split}
\end{equation}
where the coefficients in the expansion are separately 
\begin{equation}\begin{split}
    a_k &=\frac{1}{k!}x'(s)\cdot x^{(k+1)}(s),\\
    c_k & =\sum_{m=1}^{k-1}\frac{1}{m!(k-m)!}x^{(m)}(s)\cdot x^{(k-m)}(s),
\end{split}
\end{equation}
and for given any closed and smooth instanton trajectories, both Taylor expansions in the denominator and numerator are of infinite radius of convergence separately. Therefore both expansions are uniformly convergent separately, see 8.1 theorem in \cite{rudin1964}.

Then one can define the $n$-th order of $t$-expansion by the integral of rational series with type $[n,n]$
\begin{equation}\begin{split}
   & {}^n\!A_\e \coloneqq \int^\pp_{-\pp}\dif s\int^\pp_{-\pp}\dif t\; T_n(t),\\
   &  T_n(t)=\frac{\sum^n_{k=0} a_k t^k}{\e^2+\sum^n_{k=2}c_k t^k},
     \end{split}
\end{equation}
where the integrand is uniformly convergent to eq. \eqref{eq:integrand} due to the quotient law of convergent sequences and absence of poles for $\e\neq 0$. The uniform convergence can be shown by using theorem 7.9 in \cite{rudin1964}. Generally for given two uniformly convergent sequences $g_n\to g$, $f_n\to f$ and $g_n$ does not have zeroes in considering domain, then the quotient $f_n/g_n$ is uniformly convergent to $f/g$.  Namely the supremum of $|f/g-f_n/g_n|$
\begin{equation}
   \left|
    \frac{f}{g}-\frac{f_n}{g_n}
    \right|\le 
    \frac{\left|f\right|}{\left|g g_n\right|} \left|g-g_n\right|+
     \frac{\left|f-f_n\right|}{\left|g_n\right|} 
\end{equation}
approaches zero as $n\to \infty$, because both sequences are uniform convergent.
Thus the interchange of limit
and integral operations is valid in the corresponding domain (see  7.16 theorem in \cite{rudin1964}), i.e.
\begin{equation}\begin{split}
    &\lim_{n\to\infty} \int^\pp_{-\pp}\dif s\int^\pp_{-\pp}\dif t \frac{\sum^n_{k=0} a_k t^k}{\e^2+\sum^n_{k=2}c_k t^k} \\
    &=\int^\pp_{-\pp}\dif s\int^\pp_{-\pp}\dif t
     \lim_{n\to\infty} \frac{\sum^n_{k=0} a_k t^k}{\e^2+\sum^n_{k=2}c_k t^k}
\end{split}\end{equation}
which is exactly what we expect
$ \lim_{n\to\infty} {}^n\!A_\e =A_\e $.

\begin{acknowledgements}
Y.-F.W.\ is grateful to Claus Kiefer and Nick Kwidzinski (Cologne), Chao Li (Princeton) and Ziping Rao (Vienna).
H.G.\ is supported by ELI--ALPS project, co-financed by the European Union and the European Regional Development Fund No.\ GINOP-2.3.6-15-2015-00001.
Y.-F.W.\ is supported by the Bonn-Cologne Graduate School for Physics and Astronomy (BCGS).
\end{acknowledgements}

\bibliographystyle{unsrt}
\bibliography{instanton}   

\end{document}